# Three-dimensional close-to-substrate trajectories of magnetic microparticles in dynamically changing magnetic field landscapes

*Rico Huhnstock\*[1),2)], Meike Reginka[1)], Claudius Sonntag[3)], Maximilian Merkel[1),2)], Kristina Dingel[2),3)], Bernhard Sick[2),3)], Michael Vogel[1),2)], Arno Ehresmann[1),2)]*

AUTHOR ADDRESS

1) Institute of Physics and Centre for Interdisciplinary Nanostructure Science and Technology (CINSaT), University of Kassel, Heinrich-Plett-Strasse 40, D-34132 Kassel, Germany

   E-Mail: rico.huhnstock@physik.uni-kassel.de, Phone: +49 561 804-4015

2) Artificial Intelligence Methods for Experiment Design (AIM-ED), Joint Lab of Helmholtzzentrum für Materialien und Energie, Berlin (HZB) and University of Kassel, Hahn-Meitner-Platz 1, D-14109 Berlin, Germany

3) Intelligent Embedded Systems, University of Kassel, Wilhelmshöher Allee 71-73, D-34121 Kassel, Germany

ABSTRACT

The transport of magnetic nano- or microparticles in microfluidic devices using artificially designed magnetic field landscapes (MFL) is promising for the implementation of key functionalities in Lab-on-a-chip (LOC) systems. A close-to-substrate transport is hereby instrumental to use changing particle-substrate interactions upon analyte binding for analytics and diagnostics. Here, we present an essential prerequisite for such an application, namely the label-free quantitative experimental determination of the three-dimensional trajectories of superparamagnetic particles (SPP) transported by a dynamically changing MFL above a topographically flat substrate. The evaluation of the SPP sharpness within defocused video-recorded images, acquired by an optical bright-field microscope, was employed to obtain a vertical *z*-coordinate. This method applied to a prototypical transport scheme, using the static MFL of parallel-stripe domains superposed by a particular magnetic field pulse sequence, revealed a "hopping"-like motion of the magnetic particles, previously predicted by theory. Maximum vertical particle jumps of several micrometers have been observed experimentally, corroborating theoretical estimates for the particle-substrate distance. As our findings pave the way towards precise quantification of particle-substrate separations in the discussed transport system, they bear deep implications for future LOC detection schemes using only optical microscopy.

**Introduction**

Lab-on-a-chip (LOC) systems are considered a possible solution for the rising demand for fast and inexpensive medical diagnostic tools[1–3]. For practical implementation, magnetic particles (MPs) of micro- and nanoscopic sizes with various shapes and compositions are discussed as central functional components[4–7]. Their surfaces can be customized by functional chemical groups specific to bind disease markers or pathogens from a screened body fluid, for instance. As controllable actuation is possible by applying magnetic fields, MPs provide several key functionalities of LOC systems: liquid mixing[8], biomolecule capture, analyte transport and separation, and finally detection (among others)[9,10]. To transport MPs to a designated chip area, magnetic stray fields, varying on the micrometer scale, superposed with a macroscopic dynamically changing external field have been used[11–13]. The necessary magnetic stray field landscapes (MFLs) emerge either from micro-structured magnetic elements on a substrate surface[14–17] or from topographically flat magnetic thin films with a domain pattern[18–20]. In the latter case, the domains either form naturally (e.g., in magnetic garnet films)[18] or are artificially imprinted into the thin film[21,22]. An outstanding advantage of the use of MFLs is a comparably large magnetic field gradient within a small volume (over small distances), allowing for rapid MP transport[19] with local steady-state velocities of more than 100 µm/s.[23,24] It is important to note, that the motion of MPs in these transport concepts occurs close to the substrate surface. The MP motion dynamics are mainly governed by the magnetic gradient forces, the hydrodynamic drag exerted by the surrounding liquid, and the electrostatic and electrodynamic interactions with the underlying substrate[25]. If gravitation and buoyancy are neglected, the balance of magnetic, electrostatic, and van der Waals forces determines the equilibrium distance between particle and substrate, i.e., their interplay defines the position of the particles within the magnetic field landscape[23,25], which in turn influences strongly the lateral MP motion. The experimental determination of this vertical distance would therefore be, besides the two-dimensional MP trajectories, an important additional observable for the understanding of the MP motion over a flat substrate. Moreover, resolving vertical MP positions along the lateral trajectories creates an alternative access for bio-sensing applications: the electrostatic and electrodynamic interactions between MP and substrate change upon analyte binding[23] either on the particle's or on the substrate's surface when the opposing surface remains unchanged. Therefore, determining a change in the separation distance between MP and substrate upon analyte binding possesses a large potential as a bio-detection technique.

Here, we present the application of a comparably simple method and proof of concept to experimentally determine three-dimensional (3D) trajectories of superparamagnetic particles (SPP) transported within an artificially created dynamically changing MFL above a topographically flat substrate with high temporal resolution. Changes in the SPPs' sharpness in microscope images, as they move relative to the microscope focal plane[26–28], will be used in combination with a calibration procedure and lateral single particle tracking by machine learning aided software[29] to determine the MPs' 3D motion. The experimental results will be corroborated by a numerical model to estimate the separation distance change between substrate and SPP, upon changing their lateral in-plane position. The current work lays the foundation for dedicated analyte detection in LOC devices by simply observing the 3D MP motion through a microscope and for a new multiplex method of scanning material characteristics of flat or topographic magnetic surfaces, very similar to magnetic force microscopy, with many probes operated in parallel.

**Experimental concept overview**

Directed SPP transport is induced in the current work by a superposition of the MFL with weak external magnetic field pulses, resulting in a periodic transformation of the particles' potential energy. The MFL stems from a prototypical magnetic stripe domain pattern with a head-to-head (hh) and tail-to-tail (tt) magnetization configuration, engineered in an exchange-biased (EB) thin film system using ion bombardment induced magnetic patterning (IBMP)[21,30,31]. This technique allows, in general, for the fabrication of arbitrary domain patterns in EB thin films. Here, we exemplarily describe the 3D motion behavior for SPP above such a stripe pattern. The 3D tracking was achieved by implementing in- and out of focus particle image analysis in bright-field microscopy, as known from literature[26–28,32–34]. This technique has so far only seen application in the investigation of dynamic particle behavior inside confined microfluidic structures[26,28,32] or of biological processes (in this case by mostly studying fluorescent particles)[33,35]. Specific characteristics of the particle images (e.g., diffraction rings caused by Mie scattering[36], astigmatism in images using cylindrical lenses[37]) changing with $z$-position during a focus sweep partly in combination with cross-correlation based matching algorithms had been analyzed so far[28,32,38]. The determination of the SPPs' axial coordinate with respect to the microscope's focal plane has been achieved in this work by an image sharpness-based calibration procedure suitable for the studied system. This calibration has been demonstrated to work for two different types of SPPs with different optical responses. Subsequently, the "hopping" like motion behavior of SPP that is theoretically expected for the applied actuation concept[39] could be quantitatively resolved.

## Results

**Calibration procedure for the $z$-coordinate**

Superparamagnetic particles (SPP) with sizes in a diameter range between 2.8 µm and 4 µm have been placed inside a microfluidic chamber on top of a magnetically patterned and topographically flat substrate. The magnetic pattern consisted of 5 µm-wide stripe domains with a periodically alternating hh/tt magnetization direction (see arrows in Figure 1b). An additional spacer layer of Poly(methyl methacrylate) (PMMA) of 150 nm thickness has been deposited on top of the magnetic substrate to prevent SPP adsorption to the substrate surface.[11,19,23] As shown earlier, MPs are attracted to the magnetic domain walls, since they represent local minima of the MPs' potential energies[19,22,24,40,41]. Without applied external fields, the positions above the two domain wall types (hh or tt) are energetically degenerate for the SPPs. Subsequent lateral transport of SPP is achieved by applying external trapezoidal magnetic field pulses in $x$- and $z$-directions[19,22,23,42].

MP motion has been video-recorded by an optical bright-field reflection microscope with a temporal resolution of 1000 frames per second (fps). SPPs $z$-positions have been determined in the individual frames by correlating image changes caused by SPP focusing/defocusing to SPP images, which are calibrated for a $z$-position relative to the focal plane of the microscope. Hence, for the present experiments, an existing setup consisting of an optical unit and electromagnets (for applying external magnetic fields)[19] was extended by a piezo stage controlled sample holder (see Figure 1a), which moves the sample holder vertically (z-direction) with a possible technical resolution of 3 nm. This way, SPPs at their equilibrium position within the MFL have been swept through the focal plane of the used 100x magnification objective (Figure 1b). Depending on the SPP position below the focal plane, in focus, or above the focal plane, images with different signal-to-noise ratios and particle sharpness were recorded. These images are exemplarily shown for micromer®-M SPPs with a diameter of 3 μm in Figure 1c to e.

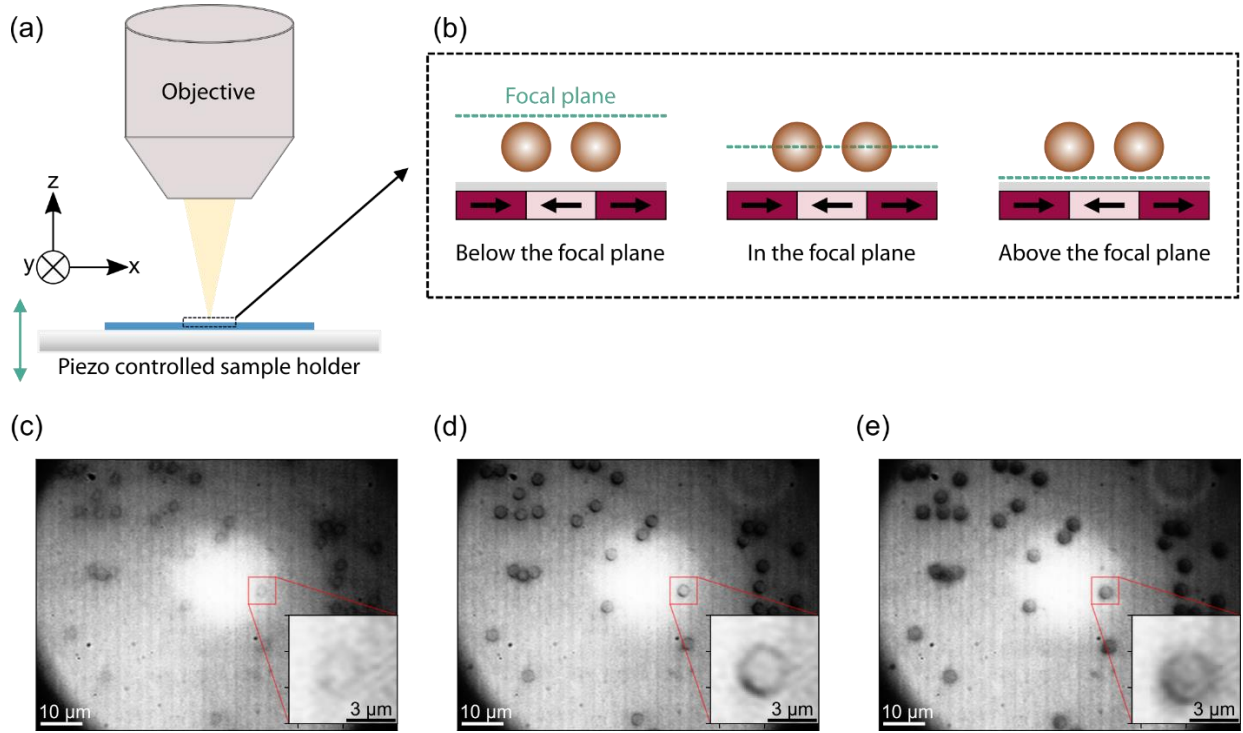

Figure 1: Acquisition of in and out of focus bright-field microscope images for axial position determination of SPPs. (a) Sketch of a focus sweep procedure. (b) SPPs are trapped within the local MFLs above a magnetic substrate with head-to-head/tail-to-tail magnetized stripe domains (indicated by black arrows). SPPs are moved from a position below the focal plane (c), through the focal plane (d) to a position above it (e). Exemplary images for micromer®-M SPP ($d$ = 3 μm) are shown for the three conditions. These images are used to calibrate the z-position of the SPP relative to the focal plane within transport experiments.

The optical axial resolution is described by the depth of field (DOF) of the utilized microscope setup. The DOF characterizes the axial range throughout an imaged specimen, where all object planes are simultaneously in focus. In the diffraction limit, the DOF can be expressed by $\text{DOF} = \frac{\lambda}{2 \cdot \text{NA}^2}$, where $\lambda$ describes the wavelength of the used light and NA is the numerical aperture of the objective[35]. The DOF of the present setup (central wavelength $\lambda \approx$ 500 nm for white light, NA = 1.4) is estimated to be approximately 130 nm. Thus, the comparably large NA not only provides the high lateral image resolution needed for two-dimensional particle tracking but also results in a rather thin focal plane which is critical for a high resolution in determining the axial SPP position with respect to that focal plane. For a quantification of the particle $z$-distance relative to the microscope's focal plane, a suitable metric for correlation of the particle image characteristics to axial coordinate during a focus sweep is needed. Approaches based on calculated point-spread functions for the object images[43], on the analysis of diffraction rings (caused by Mie scattering)[36] surrounding the object images, or on the evaluation of the intensity averaged particle radius[26] in the images are known. Instead, however, we employ a metric based on the overall intensity gradient inside an image, known as the Tenenbaum (TB) gradient[44–46], which is an established measure for the sharpness of an image object and can be used for the development of autofocusing algorithms[44,45]. In an own separate study, metrics using the TB gradient and particle edge peak profiles have been compared, where the TB gradient metric achieved the best $z$-height resolution (see Figure S2.1 in the Supporting Information). Given an image of $M \times N$ pixels², the TB gradient $f_{\text{TBG}}$ sums over the squares of all $X$ and $Y$ image gradients $G_x(i,j)$ and $G_y(i,j)$ in pixels $(i,j)$:[44]

$$f_{\text{TBG}} = \sum_i^M \sum_j^N G_x(i,j)^2 + G_y(i,j)^2.$$

$G_x(i,j)$ and $G_y(i,j)$ are convolutions of the image with Sobel operators[47], which calculate an approximation of the image intensity gradient at a given position. For visualization, the convolved images $G_x(i,j)$ and $G_y(i,j)$ for the SPP containing images in Figure 1d and 1e are shown in the Supporting Information (Figure S3.1). $f_{\text{TBG}}$ has been determined for focus sweeps of two different types of SPP: micromer®-M ($d$ = 3 µm, 4 µm) and Dynabeads™ M-270 ($d$ = 2.8 µm). While for micromer®-M SPPs the superparamagnetic material is placed around a non-magnetic organic core[48], Dynabeads™ M-270 SPPs consist of a homogenous distribution of superparamagnetic material throughout the whole particle[49]. Consequently, the optical responses of the two SPP particle types are different, and this fact is used to investigate whether the TB gradient approach works for MPs with differing image characteristics. In both cases, focus sweeps have been carried out by taking images at height steps of 25 nm (50 nm for 4 µm SPP). The images of 800 pixels x 600 pixels have been contrast enhanced and an inhomogeneous illumination background has been subtracted (for details see method section). SPPs within contrast-enhanced and background-subtracted images have been localized laterally by a machine-learning aided tracking algorithm[29] (see method section for further details). The center of the localized SPP has then been used to crop the image into regions of interest (ROIs), which were placed symmetrically around the center. The sizes of the ROIs ranged from 50 pixels x 50 pixels (corresponding to 5.5 µm x 5.5 µm) for 3 µm micromer®-M and 2.8 µm Dynabeads™ M-270, and 60 pixels x 60 pixels (corresponding to 6.6 µm x 6.6 µm) for 4 µm micromer®-M SPP, respectively.

For each obtained ROI, $f_{\text{TBG}}$ has been calculated along all recorded reference images representing a known $z$-position, which is plotted in Figure 2a for an exemplary Dynabead M-270 SPP (2.8 µm) and in Figure 2b for an exemplary micromer®-M 3 µm SPP. Both data sets show a maximum for the TB gradient, which determines the $z$-position of the highest particle sharpness (see ROI of a particle in the insets of Figure 2 plots). With $z$-positions above and below this reference point, the TB gradient decreases for both types of SPP. The data also indicates that for very high $z$-distances between particle position and focal plane, the TB gradient reaches a saturation state. This observation agrees well with previously reported investigations, prominently to be found in studies on autofocusing algorithms[44,45]. The present data, however, does not reflect a symmetric, Gaussian-like distribution of the TB gradient around the in focus position. For both particle types, the increasing flank is steeper than the decreasing one.

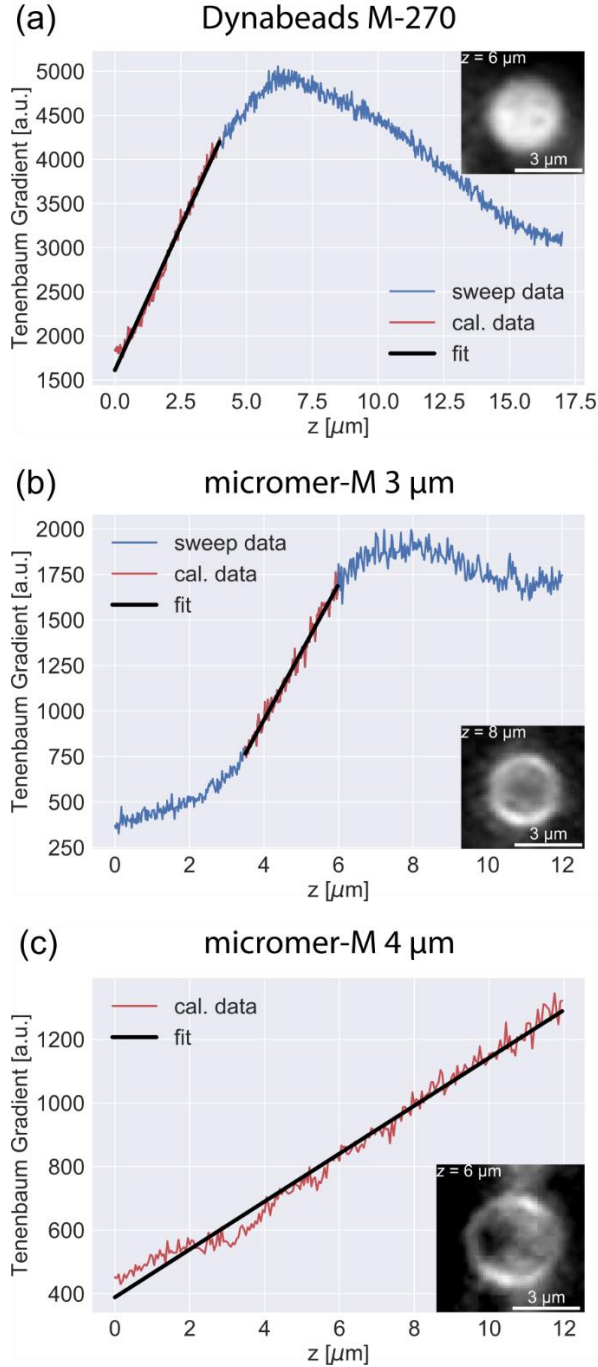

Figure 2: Exemplary calibration data for determining the *z*-position of SPPs within a microfluidic transport system. Curves (a-c) show the TB gradient, as a measure for particle sharpness, in dependence on the set *z*-position of the whole sample. The data was acquired by sweeping the sample through the focal plane of the used microscope in 25 nm steps for Dynabeads™ M-270/micromer®-M 3 µm, and in 50 nm steps for micromer®-M 4 µm, recording reference images at every step. The region of highest sensitivity (steepest slope) for *z*-position detection is highlighted in red for which a linear function can be used to fit the calibration data (black line). For micromer®-M 4 µm SPP (c), only the linear regime of the highest sensitivity calibration data is displayed which was taken for further experiments. Exemplary microscopic images of the respective SPP at the indicated *z*-position are shown in the insets of each plot.

Maximum sensitivity regarding changes in *z*-position is achieved by using the increasing branches of the calibration data. In this regime (highlighted in red in Figure 2a,b), the TB gradient changes almost linearly with increasing *z*-position in a height range of about 2 – 4 µm. Hence, a linear function has been fitted to this data (black lines in Figure 2a,b). The inversed slope of these functions amounts to $0.001 \frac{\mu m}{\text{TBG unit}}$ for Dynabeads™ M-270 and $0.003 \frac{\mu m}{\text{TBG unit}}$ for micromer®-M 3 µm. Consequently, particles have been placed at positions below the focal plane for the 3D analysis of the SPP transport motion in the following experiments. For micromer®-M 4 µm SPPs, which were investigated in the transport experiment, the TB gradient focus sweep data is displayed in Figure 2c for the linearly behaving regime with changing *z*-position. The linear fit allows direct quantification of the *z*-position with respect to the focal plane when $f_{\text{TBG}}$ has been determined from the individual frames of the transport experiment. For micromer®-M 4 µm particles the inversed linear slope of the fit function has been determined to be $0.014 \frac{\mu m}{\text{TBG unit}}$.

**3D analysis of SPP transport**

Lateral transport of SPPs in the MFL over a magnetically stripe patterned substrate is inducible by periodic external magnetic field pulses in *x*- and *z*-directions[19,22,42]. The superposition of the static local stray fields and the external magnetic field transforms the potential energy landscape of SPPs in close vicinity to the substrate in such a way, that a stepwise movement can be forced[19]. In the present work, the behavior of micromer®-M SPP (*d* = 4 µm) has been studied. One transport step towards the adjacent domain wall is inducible[19] upon a switch of the external *z*-field from $+H_{ext,z,max}$ to $-H_{ext,z,max}$. This is schematically shown in Figure 3a together with the expected 3D movement of an SPP: After a sign change in $H_{ext,z}$, the magnetic force on the SPP is altered from attractive to repulsive, initially increasing the elevation of the SPP above the substrate.

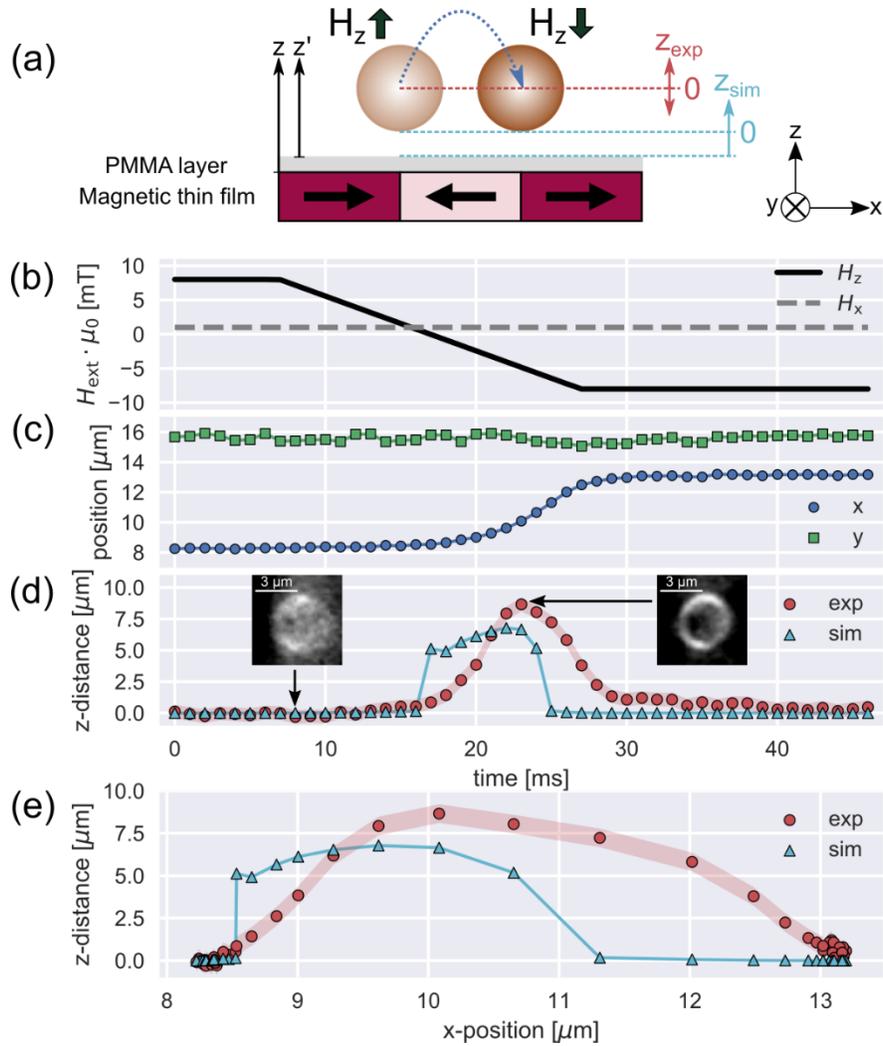

Figure 3: 3D analysis of transport dynamics for 4 μm sized SPPs within a time-dependent MFL. (a) Sketch of one transport step for the SPP above a hh/tt magnetic stripe pattern (black arrows) after changing the external magnetic field in z-direction. The blue arrow indicates the expected "hopping" like motion of the SPP. Reference positions for the normalization of the experimentally determined vertical movement ($z_{exp}$) and the theoretically estimated equilibrium distance ($z_{sim}$) are indicated. Relevant forces for the estimation of $z_{sim}$ were calculated along the indicated vertical axes z and z' (see Supporting Information S5). (b) Applied external magnetic fields as a function of time for one pulse sequence. Black solid line: field in z-direction, grey dashed line: field in x-direction. (c) Plot of an exemplary SPP trajectory: determined x- (blue circles) and y- (green squares) SPP-center coordinates as functions of time. (d) Experimentally determined z-SPP center coordinates, relative to the equilibrium position of the SPP before the z-field change, superposed on the lateral movement (red circles). The initial z-position before the z-field change has been normalized to zero. Red shaded areas indicate the fit uncertainty for the used z-coordinate calibration function. For comparison, theoretical equilibrium distances between SPP and the substrate surface were computed (cyan triangles). The insets show background-subtracted microscopic images of the SPP at the indicated time. Changes in SPP sharpness and thereby z-position are observable. (e) Plot of determined z-positions in dependence of the x-positions for an SPP during a transport step. Again, red circles represent normalized experimental data (with the red shaded area as the uncertainty), while the cyan triangles depict simulated equilibrium distances. The data normalization was performed as described for (d). Data points of x-, y-position, and the simulated z-distance were connected to provide a guide to the eye.

As the SPP moves laterally in *x*-direction towards the new location of minimum potential energy, magnetic attraction increases, thus, forcing the SPP back to its height above the substrate at the beginning of the transport step. For a test of the current 3D tracking strategy and to verify the true 3D particle motion experimentally, an external magnetic fields sequence was applied as shown in Figure 3b. The field in *x*-direction has been kept constant at $H_{\text{ext,x,max}} \cdot \mu_0 = 1$ mT. The *z*-field has been changed linearly from $H_{\text{ext,z,max}} \cdot \mu_0 = +8$ mT to $H_{\text{ext,z,max}} \cdot \mu_0 = -8$ mT within 20 ms. This field change results in *x*- and *y*-trajectories, as presented for an exemplary SPP in Figure 3c. Lateral *x*- and *y*-coordinates have been determined with the same tracking method[29] employed for particle position evaluation during focus sweeping. As expected, the *z*-field change leads to a step-like motion along the *x*-coordinate (blue circles in Figure 3c) with the step width resembling the domain width of the magnetically patterned substrate. As the MFL does not change along *y*, the *y*-position (green squares in Figure 3c) remains approximately constant apart from a small Brownian motion[19].

From the microscope images, taken with a frame rate of 1000 fps, already qualitatively an axial *z*-movement can be identified during lateral SPP motion. This is illustrated by the microscope snapshots of an exemplary SPP in the insets of Figure 3d. Prior to a *z*-field change, the SPP image shows low sharpness (and therefore a low $f_{\text{TBG}}$), since the sample was placed below the microscope focal plane by the piezo stage. After the start of the lateral *x*-movement, a significant increase in sharpness/$f_{\text{TBG}}$ is observable. This corresponds to an upward movement of the SPP, as it is gradually elevated towards the focal plane of the microscope. For the quantitative analysis of the *z*-position, $f_{\text{TBG}}$ was calculated for the cropped SPP in each recorded, background-subtracted frame and then turned into a *z*-coordinate using the calibration data. For the SPP in Figure 3d (red circles), the particle is elevated up to approximately 8.7 µm above the initial *z*-position (taken as 0 for reference) upon an inversion of the external magnetic field in *z*-direction. As the SPP completes its lateral movement along *x*, the *z*-position decreases again back to the initial *z*-elevation.

**Discussion**

The qualitative progression of the particle's motion strongly resembles the theoretically predicted "hopping" like behavior[39] (see Figure 3a), which is underpinned by the tracked *z*-position in dependence on the *x*-position in Figure 3e (red circles). Here, with the SPP moving from about 8 µm to about 13 µm, the $z(x)$-data exhibits an asymmetry with respect to the position of maximum elevation away from the substrate. The initial increase of the *z*-distance seems to be steeper than the following decrease back to the reference position. To compare the measured $z(x)$ of SPPs within the lateral transport step with theoretical estimates, SPP-substrate distances have been calculated based on the equilibria of acting forces. Since the corresponding model is extensively described in previous works[23,25], it will only be summarized briefly in the following, and the relevant equations and parameters will be given in the Supporting Information S5.

Attractive and repulsive forces, acting on the SPP along the *z*-axis, have been balanced for the calculation of an equilibrium distance between SPP and substrate at each time step (1 ms) during the transport experiment. The forces taken into account are the magnetic force $F_\text{M}(z)$, the electrostatic force $F_\text{el}(z')$, the van der Waals force $F_\text{vdW}(z')$, and the effective gravitational force $F_\text{Grav}$ (considering gravity and buoyancy simultaneously)[25]. As $F_\text{M}$ for a SPP is determined by the derivative of its potential energy $U_\text{p}$ within the MFL and consequently by the gradient of the effective magnetic field $\vec{H}_\text{eff} = \vec{H}_\text{ext} + \vec{H}_\text{MFL}$ [25], the superposition of the local static MFL $\vec{H}_\text{MFL}$ and the externally applied field $\vec{H}_\text{ext}$ needs to be calculated[23]. The *x*- and *z*-components of the static MFL $H_{\text{MFL},x}(z,x)$ and $H_{\text{MFL},z}(z,x)$ estimated from simulations are plotted in the Supporting Information S5. In previous studies[23,24], the equilibrium distance between magnetic particle and substrate surface has been calculated for a position over a domain wall center in the underlying magnetic thin film system. For estimating the change of this distance upon lateral particle motion from one domain wall to the adjacent one, $H_\text{eff}$ needs to be computed as a function of the *x*-coordinate. This has been done by simulating the MFL of a parallel-stripe domain pattern, where $H_\text{MFL}$ has been derived for different *z*- and *x*-positions[23] from results of micromagnetic simulations using the object oriented micromagnetic framework OOMMF[50]. $H_{\text{MFL},z}(x,z)$ and $H_{\text{MFL},x}(x,z)$ were subsequently approximated by an exponential decay function and a sixth-degree polynomial, respectively. The experimentally observed *x*-trajectory $x(t_i)$ of the SPP has been used to determine $F_\text{M}(z(x(t_i)))$ at each time step $t_i$ based on the effective magnetic field $H_{\text{MFL},z}(x(t_i)) + H_{\text{ext},z}(t_i)$. Using this magnetic force, the force equilibrium leads to a corresponding equilibrium distance between SPP and substrate surface, which has been calculated as functions of $t_i$ and $x(t_i)$. The results are displayed as cyan triangles in Figure 3d and e, respectively, with the smallest calculated equilibrium distance set to zero for comparison. The presented differences in the equilibrium distances are qualitatively comparable to the progression of the experimentally determined $z$-positions. In the calculations, the equilibrium distance increase and subsequent decrease appear at earlier times/smaller $x$-positions. The temporal increase and decrease of calculated equilibrium distances, moreover, emerges more abruptly compared to the experimental data. We stress that the estimated $x$-position-dependent equilibrium distances between substrate and SPP cannot fully reproduce the experimentally observed dynamic vertical motion, as the calculations do not consider inertial forces in the viscous medium due to friction[39]. Thus, for the chosen theoretical estimation of $z$-motion, the SPP jumps immediately upwards to the newly calculated equilibrium distance, as the previously attractive magnetic force $F_M(z)$ is transformed into a repulsive force after the induced change in the external $z$-field. The differences between the present static force equilibrium model and reality will be more prominent for faster field switching. Figure 3 vividly shows the delayed $z$-motion due to the inertial force, but also demonstrates that the pulses used in the experiment are long enough that the equilibrium $z$-elevation is reached by the SPP in one transport step. The estimate predicts a maximum SPP elevation difference within one transport step of 6.8 µm, which is close to the experimentally determined maximum value of $(8.7 \pm 0.5)$ µm. For the investigated 4 µm micromer®-M SPP, an uncertainty for the experimentally determined relative $z$-elevations of approximately 500 nm has been obtained (see Supporting Information S4 for further details). This uncertainty is indicated in Figures 3d and 3e (shaded areas).

The results presented in Figure 3 have been obtained for one exemplary SPP. Within the same experiment, the $z$-movement of 4 additional particles could be evaluated reliably (trajectories are shown in Supporting Information S6). Averaging over all maximum height jumps, a value of $(6 \pm 2)$ µm has been obtained, being quantitatively comparable to the theoretically estimated maximum $z$-elevation of 6.8 µm. Here it should be noted that the used model tends to overestimate the acting magnetic force[23], possibly resulting in overestimated $z$-distance changes. Differences in experimentally determined SPP $z$-elevations may be due to variations in the local MFL, the thickness of the deposited PMMA layer, and differences in the actual size and susceptibility of the investigated SPP.

## Conclusion

In this work, we presented a fully quantitative 3D analysis of SPP trajectories moved by dynamically transformed MFL. These MFL emerge from a prototypical exchange-biased magnetic thin film substrate with engineered 5 µm wide magnetic head-to-head/tail-to-tail parallel-stripe domains. A surface-near lateral transport of SPP in a microfluidic environment was induced by the superposition with time-dependent external magnetic fields. The non-trivial quantitative determination of the axial $z$-movement has been achieved by analyzing the particle sharpness in microscope images when the SPP are observed at different distances to the focal plane in combination with the automatic acquisition of reference images during a focus sweep. The TB gradient for cropped SPP images as a measure of image sharpness was shown to be a powerful metric for $z$-coordinate determination in this type of experiment. A specific TB gradient, therefore, correlates to a defined $z$-position of the SPP. The validity and applicability of this approach have been demonstrated for two different types of SPP with different optical responses: micromer®-M and Dynabeads™ M-270. With this technique, a frame-by-frame evaluation of 1000 fps videos from SPPs moving on 3D trajectories was possible. The theoretically expected "hopping" like motion of SPPs, crossing a stripe domain to reach the adjacent potential energy minimum above a neighboring domain wall, could be proven unequivocally. The maximum average axial elevation of the investigated SPP was quantified to be $(6 \pm 2)$ µm for micromer®-M SPP with $d = 4$ µm, as compared to a theoretically estimated value of around 6.8 µm. The findings pave the way for an in-depth understanding of magnetic particle dynamics within tailored magnetic stray fields, especially for application in LOC systems. The presented method is label-free and capable to be applied in high as well as standard frame rate video sequences. The quantifiable movement along the $z$-axis will open a new way for the detection of analytes in diagnostic LOC devices by optical bright-field microscopy and for material analysis of surfaces covered by liquids: as the interactions between SPP and surface will be modified upon either particle surface or substrate surface coverage with analytes, the equilibrium distances between SPP and substrate will change accordingly. With the presented method, these changes can be quantified, indicating the presence of an analyte. Similarly, surface material differences, local varieties in the MFL, or even more general differences in the acting forces may be quantified, possibly observing many particle heights and trajectories simultaneously.

## Methods

**Fabrication of magnetically patterned substrate**

The prototypically used magnetic stripe domain pattern was imprinted into an exchange-biased magnetic thin film system by applying IBMP[21,30,31]. The thin film system consists of a $Cu^{10\ nm}/Ir_{17}Mn_{83}^{30\ nm}/Co_{70}Fe_{30}^{10\ nm}/Au^{10\ nm}$ layer system deposited onto naturally oxidized Si (100) via rf-sputtering at room temperature. Field cooling initialized the in-plane direction of the EB by annealing in a vacuum chamber (base pressure = $5 \times 10^{-7}$ mbar) at 300 °C for 60 min in an in-plane magnetic field of 145 mT. For IBMP, a photoresist, that was thick enough to prevent 10 keV He ions from penetrating the magnetic layer system, was spin-coated onto the sample and 5 µm wide stripe structures (periodicity of 10 µm) perpendicular to the initial EB direction were fabricated by photolithography (Karl Suss MA-4 Mask Aligner). Subsequently, a home-built Penning ion source was used to bombard the sample with He ions at a kinetic energy of 10 keV (ion dose = $1 \times 10^{15}$ cm$^{-2}$). To obtain a hh/tt magnetic domain configuration within the thin film system, a homogenous magnetic field (100 mT), antiparallel to the initial EB direction, was applied in-plane during ion bombardment. After bombardment, removal of the photoresist was achieved by treating the sample successively in an ultrasonic bath for 5 min at 50 °C in a 3% KOH solution and for 3 min at 50 °C in acetone and water. The sample was cleaned with acetone, isopropanol, and water and dried in a $N_2$ stream. Finally, a 150 nm thick PMMA layer was deposited on top of the sample by spin-coating.

**Focus sweeps and particle transport**

For the calibration of the *z*-coordinate for SPP, a transport substrate with SPP residing on top was swept in discrete steps through the focus plane of the used microscope. Investigated SPP were obtained from micromod Partikeltechnologie GmbH (micromer®-M) and Thermo Fisher Scientific Inc. (Dynabeads™ M-270 Carboxylic Acid). Beforehand, 20 µl of a diluted SPP dispersion in water was pipetted into a microfluidic chamber on top of the transport substrate. The chamber was fabricated by cutting a ca. 5 mm x 5 mm sized window into a square sheet of Parafilm® and attaching the resulting structure to the substrate. After SPP deposition, the chamber was sealed with a cover glass (00 strength) and a drop of immersion oil (AppliChem GmbH, A0699,0100) was placed onto it. Subsequently, calibration reference images (800 pixels x 600 pixels) were recorded by using an optical bright-field microscope with a 100x magnification objective (Nikon, N.A. = 1.4) and a high-speed camera (Optronis CR450x2). For this purpose, the sample was moved axially in 25 nm steps (3 µm SPP) or 50 nm steps (4 µm SPP) via a Piezo stage (P-603.3S2 connected to PI E-625 PZT Servo Controller) and after each step, an image was taken. Directly after the sweep, transport experiments were performed using a home-built setup consisting of electromagnets for the application of homogenous magnetic fields in *x*- and *z*-direction. The sample had been placed into the setup with its plane being parallel to the direction of the magnetic field generated by the *x*-coils. Hence, the magnetic field in *z*-direction was perpendicular to the substrate plane. The magnetic stripe domains long axis was aligned perpendicular to the *x*-direction. Controlled particle motion was induced by applying a time-dependent field in the *z*-direction as sketched in Figure 3b. Videos of the particles' motion from one to the next domain wall were recorded with the same camera using a 1000 fps frame rate.

**Particle tracking**

Lateral SPP position in *x*- and *y*-direction was tracked using the Python-based program AdaPT, which is specifically designed for the analysis of spherical magnetic particles[29]. Particle locations are hereby identified in each frame using an intensity-based method. Machine learning techniques are employed to derive optimal parameters for the localization algorithm. Subsequently, particle trajectories are obtained via a frame-by-frame linking procedure. The found *x*- and *y*-positions of the SPP are used to isolate each tracked particle for the *z*-coordinate measurements. Prior to that, raw images were contrast-enhanced through histogram equalization and background noise was reduced by applying low-pass filtering to the Fourier transform of the images with a following inverse Fourier transformation conducted. A subsequent subtraction of the inhomogeneous illumination background was done by averaging over all images of the SPP transport videos and then afterwards subtracting the obtained average image from all background-including images (focus sweep and transport). The background subtraction is a required preprocessing step for the images from subsequently performed SPP transport experiments: When particles change their lateral position, they can be exposed to different lighting conditions along their trajectory. Here, the background subtraction minimizes the illumination dependent changes of $f_{TBG}$ which finally allows using the calibration data. In the processed images, ROIs sized 50 pixels x 50 pixels (3 µm SPP) and 60 pixels x 60 pixels (4 µm SPP) were cropped around each SPP, with the particle being centered. The TB gradient was computed for the ROI in each time frame of the recorded videos and *z*-coordinate identification was completed by using the respective calibration function from the focus sweep data.

ASSOCIATED CONTENT

**Supporting Information**.

S1 Image processing

S2 Comparison of metrics

S3 Application of Sobel operator

S4 Estimation of z-coordinate uncertainty

S5 Theoretical model: important equations & parameters

S6 Vertical trajectories for all evaluable particles

AUTHOR INFORMATION

**Corresponding Author**

*rico.huhnstock@physik.uni-kassel.de

**Author Contributions**

RH: Investigation, Formal analysis, Writing – original draft

MR: Formal analysis, Writing – review & editing

CS: Formal analysis, Software

MM: Formal analysis, Writing – review & editing

KD: Software, Writing - review & editing

BS: Supervision, Writing – review & editing

MV: Supervision, Project administration, Writing – review & editing

AE: Supervision, Conceptualization, Project administration, Funding acquisition, Writing – review & editing

**Funding Sources**

---

ACKNOWLEDGMENT

The authors thank the Center for Interdisciplinary Nanostructure Science and Technology (CINSaT) and the AIM-ED Joint Lab at the University of Kassel for promoting cross-disciplinary communication and research, as well as project "MASH" supported by an internal grant from the University of Kassel. Further, we thank Dr. Dennis Holzinger for fruitful input in the conceptualization phase of this project.

ABBREVIATIONS

LOC = Lab-on-a-chip

MP = magnetic particle

MFL = magnetic field landscape

3D = three-dimensional

SPP = superparamagnetic particle

hh = head-to-head

tt = tail-to-tail

EB = exchange-bias

IBMP = ion bombardment induced magnetic patterning

PMMA = Poly(methyl methacrylate)

DOF = depth of field

TB gradient = Tenenbaum gradient

ROI = region of interest

REFERENCES

1.	Knight, J. Honey, I shrunk the lab. *Nature* **418**, 474–475 (2002).

2. Kricka, L. J. Microchips, microarrays, biochips and nanochips: Personal laboratories for the 21st century. in *Clinica Chimica Acta* **307**, 219–223 (Elsevier, 2001).

3. Holland, C. A. & Kiechle, F. L. Point-of-care molecular diagnostic systems - Past, present and future. *Current Opinion in Microbiology* **8**, 504–509 (2005).

4. Gijs, M. A. M. Magnetic bead handling on-chip: New opportunities for analytical applications. *Microfluidics and Nanofluidics* **1**, 22–40 (2004).

5. Pankhurst, Q. A., Connolly, J., Jones, S. K. & Dobson, J. Applications of magnetic nanoparticles in biomedicine. *J. Phys. D. Appl. Phys.* **36**, R167–R181 (2003).

6. Pamme, N. Magnetism and microfluidics. *Lab on a Chip* **6**, 24–38 (2006).

7. Ruffert, C. Magnetic Bead—Magic Bullet. *Micromachines* **7**, 21 (2016).

8. Holzinger, D. & Ehresmann, A. Diffusion enhancement in a laminar flow liquid by near-surface transport of superparamagnetic bead rows. *Microfluid. Nanofluidics* **19**, 395–402 (2015).

9. Abedini-nassab, R., Pouryosef Miandoab, M. & Şaşmaz, M. Microfluidic synthesis, control, and sensing of magnetic nanoparticles: A review. *Micromachines* **12**, 768 (2021).

10. Khizar, S. *et al.* Magnetic nanoparticles in microfluidic and sensing: From transport to detection. *Electrophoresis* **41**, 1206–1224 (2020).

11. Yellen, B. B. *et al.* Traveling wave magnetophoresis for high resolution chip based separations. *Lab Chip* **7**, 1681–1688 (2007).

12. Ennen, I. *et al.* Manipulation of magnetic nanoparticles by the strayfield of magnetically patterned ferromagnetic layers. *J. Appl. Phys.* **102**, 013910 (2007).

13. Rampini, S., Li, P. & Lee, G. U. Micromagnet arrays enable precise manipulation of individual biological analyte–superparamagnetic bead complexes for separation and sensing. *Lab Chip* **16**, 3645–3663 (2016).

14. Chen, A., Byvank, T., Vieira, G. B. & Sooryakumar, R. Magnetic microstructures for control of brownian motion and microparticle transport. *IEEE Trans. Magn.* **49**, 300–308 (2013).

15. Sarella, A., Torti, A., Donolato, M., Pancaldi, M. & Vavassori, P. Two-dimensional programmable manipulation of magnetic nanoparticles on-chip. *Adv. Mater.* **26**, 2384–2390 (2014).

16. Sajjad, U., Lage, E. & McCord, J. A Trisymmetric Magnetic Microchip Surface for Free and Two-Way Directional Movement of Magnetic Microbeads. *Adv. Mater. Interfaces* **5**, 1801201 (2018).

17. Rampini, S. *et al.* Design of micromagnetic arrays for on-chip separation of superparamagnetic bead aggregates and detection of a model protein and double-stranded DNA analytes. *Sci. Rep.* **11**, 5302 (2021).

18. Tierno, P., Sagués, F., Johansen, T. H. & Fischer, T. M. Colloidal transport on magnetic garnet films. *Phys. Chem. Chem. Phys.* **11**, 9615–9625 (2009).

19. Holzinger, D., Koch, I., Burgard, S. & Ehresmann, A. Directed Magnetic Particle Transport above Artificial Magnetic Domains Due to Dynamic Magnetic Potential Energy Landscape Transformation. *ACS Nano* **9**, 7323–7331 (2015).

20. Ehresmann, A. *et al.* Asymmetric magnetization reversal of stripe-patterned exchange bias layer systems for controlled magnetic particle transport. *Adv. Mater.* **23**, 5568–5573 (2011).

21. Mougin, A. *et al.* Magnetic micropatterning of FeNi/FeMn exchange bias bilayers by ion irradiation. *J. Appl. Phys.* **89**, 6606–6608 (2001).

22. Ehresmann, A., Koch, I. & Holzinger, D. Manipulation of superparamagnetic beads on patterned exchange-bias layer systems for biosensing applications. *Sensors (Switzerland)* **15**, 28854–28888 (2015).

23. Reginka, M. *et al.* Transport Efficiency of Biofunctionalized Magnetic Particles Tailored by Surfactant Concentration. *Langmuir* **37**, 8498–8507 (2021).

24. Huhnstock, R. *et al.* Translatory and rotatory motion of exchange-bias capped Janus particles controlled by dynamic magnetic field landscapes. *Sci. Rep.* **11**, 21794 (2021).

25. Wirix-Speetjens, R., Fyen, W., Xu, K., De Boeck, J. & Borghs, G. A force study of on-chip magnetic particle transport based on tapered conductors. in *IEEE Transactions on Magnetics* **41**, 4128–4133 (2005).

26. Dettmer, S. L., Keyser, U. F. & Pagliara, S. Local characterization of hindered Brownian motion by using digital video microscopy and 3D particle tracking. *Rev. Sci. Instrum.* **85**, 23708 (2014).

27. Zhang, Z. & Menq, C. H. Three-dimensional particle tracking with subnanometer resolution using off-focus images. *Appl. Opt.* **47**, 2361–2370 (2008).

28. Tasadduq, B. *et al.* Three-dimensional particle tracking in microfluidic channel flow using in and out of focus diffraction. *Flow Meas. Instrum.* **45**, 218–224 (2015).

29. Dingel, K., Huhnstock, R., Knie, A., Ehresmann, A. & Sick, B. AdaPT: Adaptable Particle Tracking for spherical microparticles in lab on chip systems. *Comput. Phys. Commun.* **262**, 107859 (2021).

30. Holzinger, D. *et al.* Tailored domain wall charges by individually set in-plane magnetic domains for magnetic field landscape design. *J. Appl. Phys.* **114**, 013908 (2013).

31. Gaul, A. *et al.* Engineered magnetic domain textures in exchange bias bilayer systems. *J. Appl. Phys.* **120**, 033902 (2016).

32. Barnkob, R., Kähler, C. J. & Rossi, M. General defocusing particle tracking. *Lab Chip* **15**, 3556–3560 (2015).

33. Zhong, Y. & Wang, G. Three-Dimensional Single Particle Tracking and Its Applications in Confined Environments. *Annual Review of Analytical Chemistry* **13**, 381–403 (2020).

34. Barnkob, R. & Rossi, M. General defocusing particle tracking: fundamentals and uncertainty assessment. *Exp. Fluids* **61**, 110 (2020).

35. Zhou, Y., Handley, M., Carles, G. & Harvey, A. R. Advances in 3D single particle localization microscopy. *APL Photonics* **4**, 060901 (2019).

36. Kovari, D. T., Dunlap, D., Weeks, E. R. & Finzi, L. Model-free 3D localization with precision estimates for brightfield-imaged particles. *Opt. Express* **27**, 29875 (2019).

37. Kao, H. P. & Verkman, A. S. Tracking of single fluorescent particles in three dimensions: use of cylindrical optics to encode particle position. *Biophys. J.* **67**, 1291–1300 (1994).

38. Taute, K. M., Gude, S., Tans, S. J. & Shimizu, T. S. High-throughput 3D tracking of bacteria on a standard phase contrast microscope. *Nat. Commun.* **6**, 8776 (2015).

39. Klingbeil, F. *et al.* Evaluating and forecasting movement patterns of magnetically driven microbeads in complex geometries. *Sci. Rep.* **10**, 1–12 (2020).

40. Koch, I. *et al.* Smart Surfaces: Magnetically Switchable Light Diffraction through Actuation of Superparamagnetic Plate-Like Microrods by Dynamic Magnetic Stray Field Landscapes. *Adv. Opt. Mater.* **6**, 1800133 (2018).

41. Koch, I. *et al.* 3D Arrangement of Magnetic Particles in Thin Polymer Films Assisted by Magnetically Patterned Exchange Bias Layer Systems. *Part. Part. Syst. Charact.* **38**, 2100072 (2021).

42. Ueltzhöffer, T. *et al.* Magnetically Patterned Rolled-Up Exchange Bias Tubes: A Paternoster for Superparamagnetic Beads. *ACS Nano* **10**, 8491–8498 (2016).

43. Ram, S., Prabhat, P., Chao, J., Ward, E. S. & Ober, R. J. High accuracy 3D quantum dot tracking with multifocal plane microscopy for the study of fast intracellular dynamics in live cells. *Biophys. J.* **95**, 6025–6043 (2008).

44. Santos, A. *et al.* Evaluation of autofocus functions in molecular cytogenetic analysis. *J. Microsc.* **188**, 264–272 (1997).

45. Sun, Y., Duthaler, S. & Nelson, B. J. Autofocusing algorithm selection in computer microscopy. in *2005 IEEE/RSJ International Conference on Intelligent Robots and Systems,*

*IROS* 70–76 (IEEE Computer Society, 2005).

46. Pertuz, S., Puig, D. & Garcia, M. A. Analysis of focus measure operators for shape-from-focus. *Pattern Recognit.* **46**, 1415–1432 (2013).

47. Pratt, W. K. Digital image processing. 750 (1978).

48. micromer®-M. (2021). Available at: https://www.micromod.de/.

49. Dynabeads$^{TM}$ M-270 Carboxylic Acid. (2021). Available at: https://www.thermofisher.com.

50. Donahue, M. J. & Porter, D. G. OOMMF User's Guide, Version 1.0. *Natl. Inst. Stand. Technol. Gaithersburg, MD* (1999).

# Supporting Information

## S1 Image processing

To optimize the microscope images before further evaluation, they were contrast-enhanced and noise-reduced. For contrast enhancement, the histogram equalization function from the "scikits-image" Python library was applied[1]. Subsequently, noise in the images was reduced by performing a Fast Fourier Transformation (FFT), using the respective function from the Python-based "Numpy" library[2], then applying a home-programmed low-pass filter, and finally performing an inverse FFT. The image data were normalized to exhibit a minimum pixel value of 0 and a maximum of 255 using the Python "Open-CV" library[3]. For comparison, raw image (a), contrast-enhanced image (b), and noise-reduced images (c) of an exemplary video frame showing several SPPs are shown in Figure S1.1.

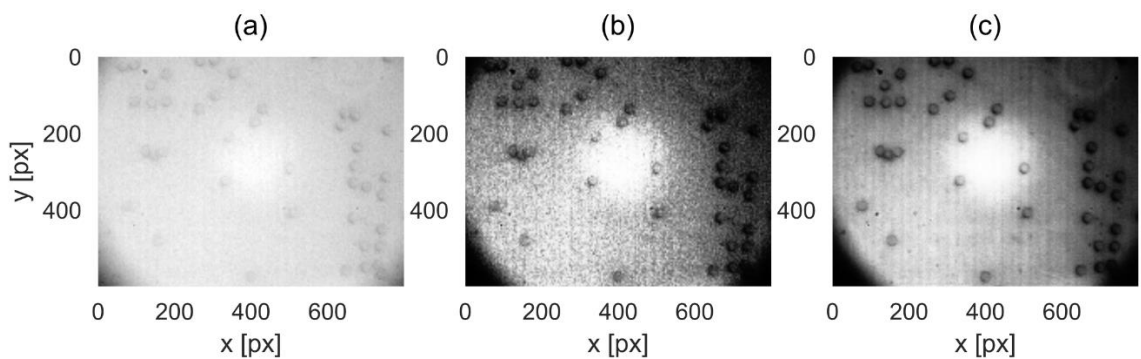

*Figure S1. 1: Display of raw (a), contrast-enhanced (b), and noise-reduced (c) images for SPP.*

## S2 Comparison of metrics

To employ the most suitable metric for the identification of SPP $z$-coordinates in the image data derived from the described experiments, the TB gradient and line profile peak heights for cropped SPP images were compared. The latter was determined by extracting a line profile across the SPP image and then calculating the peak height observable at the position of the particle edge. The results are displayed in Figure S2.1, deeming the Tenenbaum gradient more suitable, as the lower data noise provides higher $z$-coordinate accuracy.

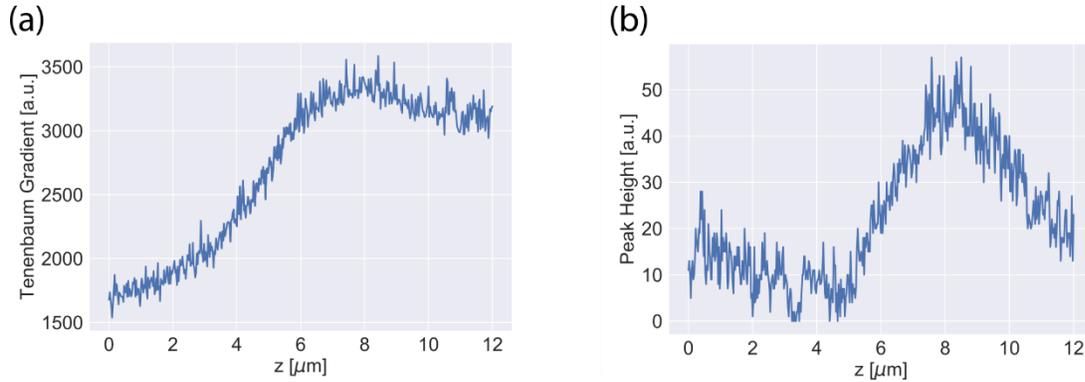

*Figure S2. 1: Plots of TB gradient (a) and line profile peak height (b) in dependence of the axial z-coordinate for non-background subtracted images of a micromer-M SPP with d = 3 µm.*

## S3 Application of Sobel operator

For the calculation of the Tenenbaum gradient, Sobel operators were applied to the optimized images in *x*-(horizontal) and *y*-direction (vertical) using the Python-based "Open-CV" library[3]. Exemplary results of these operations are shown in Figure S3.1.

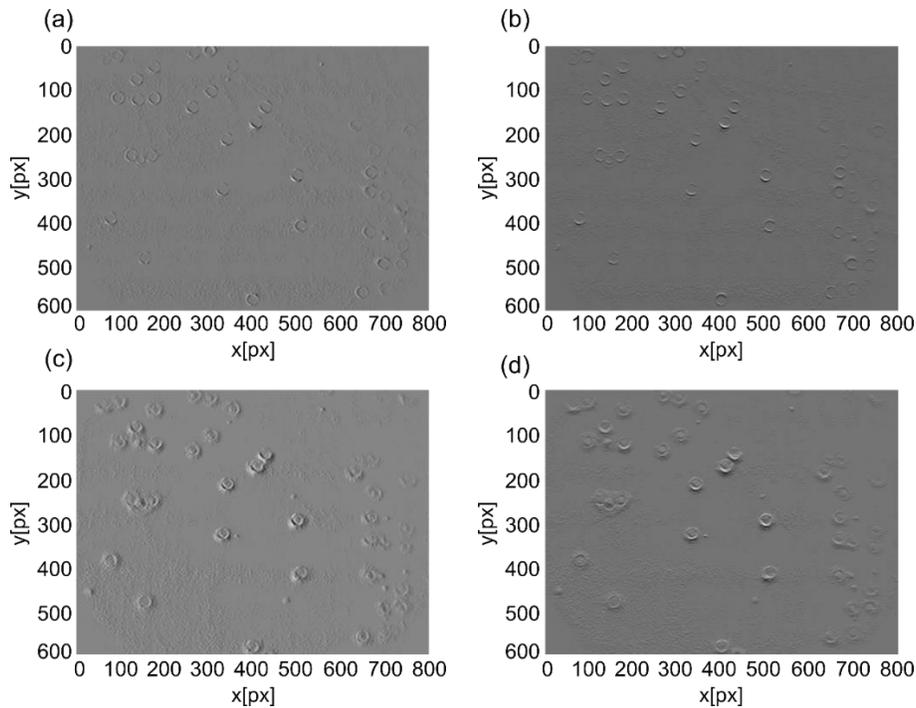

*Figure S3. 1: Convolution of optimized images with Sobel operators. (a) and (b) display the convolved images for focussed SPP in x- and y-direction, respectively, while (c) and (d) display the convolved images for SPP above the focal plane.*

## S4 Estimate of *z*-position determination uncertainty

Some experimental conditions need to be considered when assessing the presented 3D tracking of SPP in dynamically transformed MFL. First, the accuracy of the employed *z*-coordinate calibration procedure is mainly influenced by the recorded image noise and therefore the quality of the linear fit to the focus sweep data. It was found that background subtraction significantly reduces the noise in the calculated TB gradient as a function of the *z*-coordinate (see S1). Additionally, it diminishes artifacts due to inhomogeneous illumination during the experiment. The uncertainty for z-coordinate determination from the acquired fit function was calculated by considering the mean residual of experimental focus sweep data $z_{\text{sweep}}$ and linear fit $f_{\text{lin}}(F_{\text{TBG}})$:

$$\text{mean residual} = \frac{\sum_i^N \left| z_{\text{sweep,i}} - f_{\text{lin}}(F_{\text{TBG,i}}) \right|}{N},$$

with the number of sampling points *N*. Other sources of uncertainty include the accuracy of axial positioning during the focus sweep by the installed piezo stage, focus drifts during focus sweep and transport image data acquisition, and unintended movement of the sample plane due to, e.g., external vibrations. As these influences are either negligible or not quantifiable, we take the above-described measure for the error value of the determined SPP *z*-positions in the transport videos.

## S5 Theoretical model: important equations & parameters

According to the theoretical model proposed in Ref.[4], equilibrium elevations above the flat substrate for micromer©-M SPP (d = 4 µm) within a periodic magnetic field landscape (MFL) were calculated by balancing the vertically directed gravitational, buoyancy, magnetic, electrostatic, and van-der-Waals forces. The used formulae and parameters for the calculations are presented in the following:

Gravitational force:

$$F_{\text{G}} = \frac{4}{3} \cdot \pi \cdot r^3 \cdot \rho_{\text{particle}} \cdot g$$

r: particle radius

$\rho_{particle}$: particle density

g: gravitational acceleration

Buoyancy force:

$$F_B = \frac{4}{3} \cdot \pi \cdot r^3 \cdot \rho_{liquid} \cdot g$$

$\rho_{liquid}$: liquid density

Magnetic force[5]:

$$\vec{F}_m(x,z) = -\mu_0 \cdot \left(\vec{m}_b(x,z) \cdot \vec{\nabla}\right) \cdot \vec{H}_{eff}(x,z)$$

$\mu_0$: vacuum permeability

$\vec{H}_{eff}(x,z)$: superposition of external field $\vec{H}_{ext}$ and MFL $\vec{H}_{MFL}$

$\vec{m}_b(x,z)$: magnetic moment of particle calculated by Langevin function[6]:

$$\vec{m}_b(x,z) = m_s \cdot \left[\coth\left(b \cdot \vec{H}_{eff}(x,z)\right) - \left(\frac{1}{b \cdot \vec{H}_{eff}(x,z)}\right)\right]$$

$m_s$: saturation magnetic moment

$b$: Langevin parameter

Electrostatic force[7]:

$$F_{el}(z') = \frac{2 \cdot \pi \cdot \varepsilon_r \cdot \varepsilon_0 \cdot r \cdot \kappa}{1 - e^{-2\kappa z'}} \cdot \begin{bmatrix} 2 \cdot \zeta_{particle} \cdot \zeta_{substrate} \cdot e^{-\kappa z'} \\ \pm(\zeta_{particle}^2 + \zeta_{substrate}^2) \cdot e^{-2\kappa z'} \end{bmatrix}$$

$\varepsilon_r \varepsilon_0$: permittivity of the liquid

$\kappa$: inverse Debye length

$\zeta_{particle}$: particle zeta potential

$\zeta_{substrate}$: substrate zeta potential

Van-der-Waals force[8]:

$$F_{\text{vdW}}(z') = \frac{A_{\text{H},123} \cdot r}{6 \cdot z'^2}$$

$A_{\text{H},123}$: Hamaker constant for particle(1)/substrate(2)/liquid(3) system

| | |
|---|---|
| Particle radius $r$ | 2 μm |
| Particle density $\rho_{\text{particle}}$[9] | 1300 kg/m³ |
| Particle saturation magnetic moment $m_s$[5] | $4.48 \cdot 10^{-14}$ Am² |
| Langevin parameter $b$[5] | $1.05 \cdot 10^{-4}$ m/A |
| Particle zeta potential $\zeta_{\text{particle}}$[9] | -32 mV |
| Substrate zeta potential $\zeta_{\text{substrate}}$[10] | -65 mV |
| Thickness resist layer $t_{\text{resist}}$ | 150 nm |
| Thickness capping layer $t_{\text{cap}}$ | 10 nm |
| Debye length $1/\kappa$[11] | 100 nm |
| Hamaker constant $A_{\text{H},123}$[12] | $1.23 \cdot 10^{-20}$ J |
| Liquid density $\rho_{\text{liquid}}$ | 1000 kg/m³ |

*Table S5. 1: Used parameters for the calculation of relevant forces determining the steady-state distance between the particle and the underlying substrate.*

The MFL was computed by simulating the employed hh/tt magnetization pattern within the ferromagnetic $Co_{70}Fe_{30}$ layer, using the object-oriented micromagnetic framework (OOMMF)[13]. Here, a grid of $x$ = 20 μm and $y$ = 10 μm and a volume of (5 nm)³ for the cubic mesh elements within this grid was implemented. The EB related unidirectional anisotropy within each stripe domain was considered as a fixed Zeeman term with an alternating sign for differently magnetized stripes. The magnitude of the Zeeman term was chosen to be the EB field of the respective stripe domain type measured by magneto-optical Kerr magnetometry[4]. Due to the fabrication process of the domain pattern (ion bombardment induced magnetic patterning), saturation magnetization[14] and uniaxial anisotropy[15] within the bombarded stripes are decreased and values for the simulation were chosen accordingly[4]. Once the micromagnetic computation has converged, the magnetic moment for each mesh element was calculated by multiplying the obtained magnetization in the x-direction with the mesh element volume. The MFL at position $\vec{r}(x, y, z)$ consisting of the individual components $\vec{H}_x(\vec{r})$ and $\vec{H}_z(\vec{r})$ was then calculated according to the dipole approximation[16]:

$$\vec{H}(\vec{r}) = \frac{1}{4\pi} \cdot \sum_i \frac{3 \cdot (\vec{R} \cdot \vec{m}_i) \cdot \vec{R}}{|\vec{R}|^5} - \frac{\vec{m}_i}{|\vec{R}|^3}.$$

$\vec{R} = \vec{r} - \vec{r}_i$ represents the distance vector between position $\vec{r}$ and dipole position (mesh element position) $\vec{r}_i$. For visualization, the z- and x-component of the MFL are plotted in Figure S5.1 in dependence on the x- and z-coordinates above the substrate surface.

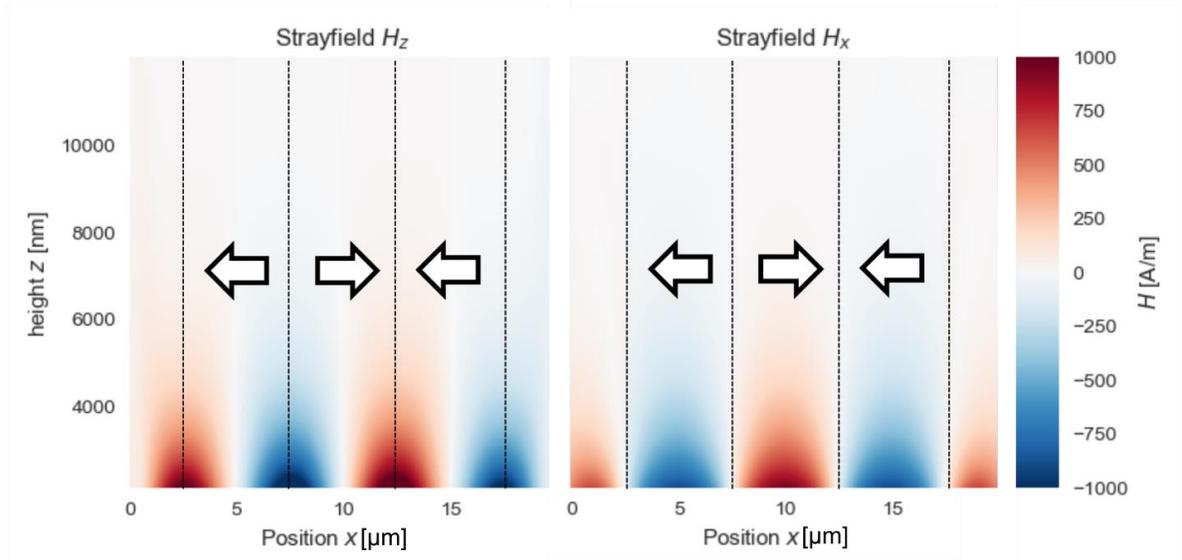

Figure S5.1: Distribution of magnetic stray fields in x- and z-direction on top of magnetically stripe domain patterned substrates with head-to-head/tail-to-tail configuration (as indicated by arrows). The magnetic field strength is color-coded. The z-dependent progression of both $H_z$ and $H_x$ results from fitting an exponential decay and a polynomial function, respectively, to distinct values of $H_{z,x}$ calculated from a magnetization pattern at a set of z-heights. The pattern was derived from micromagnetic OOMMF simulations.

S6 Vertical trajectories for all evaluable particles

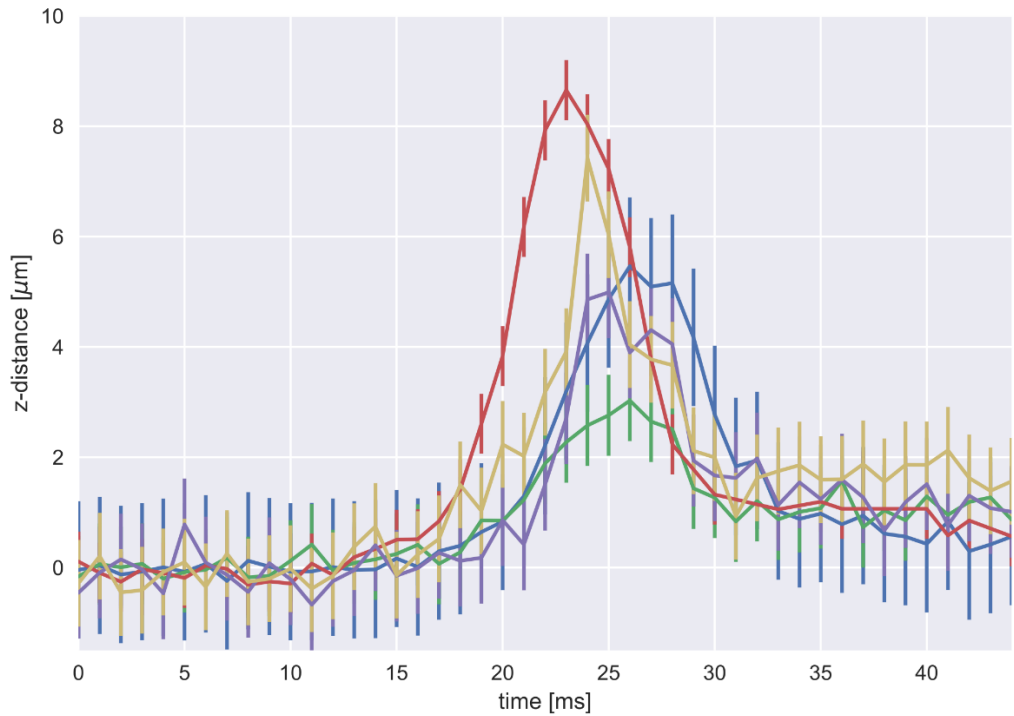

Figure S6.1: Experimentally measured vertical *z*-trajectories of five superparamagnetic particles in the analyzed particle transport experiment. Error bars are indicating the uncertainty of the respectively used calibration function for *z*. Note that all "jumps" in the displayed data might not occur at the same time, since some frames were lost during conventional two-dimensional particle position tracking.

1. Van Der Walt, S. *et al.* Scikit-image: Image processing in python. *PeerJ* **2014**, 1–18 (2014).

2. Harris, C. R. *et al.* Array programming with NumPy. *Nature* **585**, 357–362 (2020).

3. Bradski, G. The OpenCV Library. *Dr. Dobb's J. Softw. Tools* (2000).

4. Reginka, M. *et al.* Transport Efficiency of Biofunctionalized Magnetic

Particles Tailored by Surfactant Concentration. *Langmuir* **37**, 8498–8507 (2021).

5. Holzinger, D., Koch, I., Burgard, S. & Ehresmann, A. Directed Magnetic Particle Transport above Artificial Magnetic Domains Due to Dynamic Magnetic Potential Energy Landscape Transformation. *ACS Nano* **9**, 7323–7331 (2015).

6. Yoon, M. *et al.* Superparamagnetism of transition metal nanoparticles in conducting polymer film. in *Journal of Magnetism and Magnetic Materials* **272–276**, E1259–E1261 (2004).

7. Wirix-Speetjens, R., Fyen, W., Xu, K., De Boeck, J. & Borghs, G. A force study of on-chip magnetic particle transport based on tapered conductors. in *IEEE Transactions on Magnetics* **41**, 4128–4133 (2005).

8. Gregory, J. Approximate expressions for retarded van der waals interaction. *J. Colloid Interface Sci.* **83**, 138–145 (1981).

9. micromod Partikeltechnologie GmbH. Technisches Datenblatt: micromer-M, PEG-COOH, 4 µm.

10. Khademi, M., Wang, W., Reitinger, W. & Barz, D. P. J. Zeta Potential of Poly(methyl methacrylate) (PMMA) in Contact with Aqueous Electrolyte–Surfactant Solutions. *Langmuir* **33**, 10473–10482 (2017).

11. Butt, H., Graf, K. & Kappl, M. *Physics and Chemistry of Interfaces*. Wiley (2003).

12. Feldman, K., Tervoort, T., Smith, P. & Spencer, N. D. Toward a force spectroscopy of polymer surfaces. *Langmuir* **14**, 372–378 (1998).

13. Donahue, M. J. & Porter, D. G. OOMMF User's Guide, Version 1.0. *Natl. Inst. Stand. Technol. Gaithersburg, MD* (1999).

14. Huckfeldt, H. *et al.* Modification of the saturation magnetization of exchange bias thin film systems upon light-ion bombardment. *J. Phys. Condens. Matter* **29**, 125801 (2017).

15. Müglich, N. D. *et al.* Preferential weakening of rotational magnetic anisotropy by keV-He ion bombardment in polycrystalline exchange bias layer systems. *New J. Phys.* **20**, 053018 (2018).

16. Nolting, W. Grundkurs Theoretische Physik 3. Springer (2013).